\documentclass{article}
\usepackage{graphicx} 
\usepackage[text={17.1cm,24.6cm},centering]{geometry} 
\usepackage[utf8]{inputenc}
\usepackage{authblk}
\usepackage{blkarray}
\usepackage{cite}
\usepackage{comment}

\usepackage{amsfonts,amsmath,amssymb,amsthm}
\usepackage{latexsym,mathrsfs,mathtools,bm}
\usepackage{graphicx,subcaption,epsfig,caption,float,xcolor}
\usepackage{enumitem}
\usepackage{chngcntr}

\usepackage[breaklinks]{hyperref} 

\numberwithin{equation}{section}

\newcommand{\bc}{\mathbf{c}}
\newcommand{\bd}{\mathbf{d}}
\newcommand{\rr}{\boldsymbol{\rho}}
\newcommand{\orr}{\overline{\boldsymbol{\rho}}}
\newcommand{\oa}{\overline{a}}
\newcommand{\ob}{\overline{b}}
\newcommand{\oc}{\overline{c}}
\newcommand{\oN}{\overline{N}}
\newcommand{\ox}{\overline{x}}

\newcommand{\CC}{\mathbb{C}}
\newcommand{\cH}{\mathcal{H}}

\begin{document}
\title{\bf Exactly solvable inhomogeneous XY spin chain}
\author{Pierre-Antoine Bernard\textsuperscript{$1$}\footnote{E-mail: pierre-antoine.bernard@umontreal.ca}~,
Nicolas Cramp\'e\textsuperscript{$2$}\footnote{E-mail: crampe1977@gmail.com}~,
Quentin Labriet\textsuperscript{$1$}\footnote{E-mail: quentin.labriet@umontreal.ca}~,
Lucia Morey\textsuperscript{$1$}\footnote{E-mail: lucia.morey@umontreal.ca}~,
Luc Vinet\textsuperscript{$1,3$}\footnote{E-mail: luc.vinet@umontreal.ca}~
\vspace{0.2cm}\\
\textsuperscript{$1$}
\small Centre de Recherches Math\'ematiques, Universit\'e de Montr\'eal, P.O. Box 6128, \\
\small Centre-ville Station, Montr\'eal (Qu\'ebec), H3C 3J7, Canada.\vspace{0.2cm}\\
\textsuperscript{$2$}
\small CNRS -- Universit\'e de Montr\'eal CRM - CNRS,
 France.\vspace{0.2cm}\\
\textsuperscript{$3$}
\small IVADO, Montr\'eal (Qu\'ebec), H2S 3H1, Canada .\\
}
\date{}
\maketitle

\bigskip

\begin{center}
\begin{minipage}{12cm}
\begin{center}
{\bf Abstract}\\
\end{center}  
Analytical expressions for the eigenvalues of certain inhomogeneous XY spin chains are computed. These models are rewritten in terms of free-fermion models using a well-known Jordan-Wigner transformation. Finding the spectrum of such models amounts to diagonalizing a matrix whose size is equal to the number of sites in the chain. This is achieved by recognizing and exploiting contiguity relations satisfied by specific orthogonal polynomials.
\end{minipage}
\end{center}

\medskip

\section{Introduction}

Quantum spin chains describe a wide array of phenomena. They were introduced to study magnetization in metals \cite{Heisenberg1928ZurTD} and are, among other things, used to investigate phase transition and thermalization (see \cite{sachdev1999quantum, abanin2019colloquium} for general references). These endeavours are facilitated by the existence of integrable models of spin chains, which allow the exploration of these questions by exact computations. For instance, Onsager provided a rigorous demonstration of the existence of an order-disorder phase transition in the two-dimensional Ising model by exactly computing the dominant contribution to its partition function \cite{onsager1944crystal}. This has been generalized to more general statistical models in \cite{Baxter} and paved the way to the quantum inverse scattering approach, initiated by the Leningrad school \cite{FRT}.

Recently, significant attention has been devoted to analyzing the impact of introducing position-dependent couplings and magnetic fields in spin chain models. These inhomogeneities, which can be thought of as representing the effects of defects in materials, are known to significantly influence the properties of spin chains. They can lead, for example, to violations of the area law for entanglement entropy \cite{ramirez2015entanglement,vitagliano2010volume}, induce localization \cite{anderson1958absence}, affect the scaling of magnetic conductivity \cite{bernard2025currents}, and are essential for achieving perfect state transfer \cite{christandl2004perfect} and fractional revival \cite{christandl2017analytic}. An overview of some recent results is provided in \cite{bernard2024distinctive}.

As a means to investigate such phenomena, exactly solvable models of inhomogeneous XX spin chains based on discrete families of Askey–Wilson polynomials were introduced in \cite{crampe2019free} (see also \cite{CNV21,Sasaki24}). This approach relies on the three-term recurrence relation satisfied by these polynomials to diagonalize the single-particle Hamiltonian. The stationary state wavefunctions of the latter then serve to define appropriate fermionic operators, allowing the full Hamiltonian to be diagonalized, just as in the homogeneous case \cite{lieb1961two}. Closed-form expressions for the spectra and stationary states can then be obtained.
Similar constructions using multivariate orthogonal polynomials have also been used to study free-fermion models, not only on chains, but on higher dimensional lattices \cite{BERNARD2022115975} as well. Let also mention that these models are closely related to field theory and experiments using the notion of entanglement Hamiltonian \cite{Eisler22,Eisler24,Eisler244}. 

The aim of this paper is to extend the approach developed in \cite{crampe2019free} to construct exactly solvable models of XY spin chains. This involves diagonalizing a block tridiagonal matrix whose entries encode the inhomogeneous couplings and magnetic field characterizing the model. We show that, with an appropriate choice of these parameters, the block tridiagonal matrix can be diagonalized utilizing the so-called contiguity relations of Askey–Wilson polynomials. These were recently classified in \cite{Contiguity25}, where they were interpreted in terms of spectral transforms.

Section \ref{sec:XY} contains the definition of the inhomogeneous XY model as well as its connection to the free-fermion model under the Jordan--Wigner transformation. We recall that the spectrum of this model is obtained by diagonalizing a $d\times d$ matrix, where $d$ is the number of sites.
This is explored in Section \ref{sec:exact}where it is shown that this diagonalization is closely related to the $B_2$-contiguity relations which are also defined in this section.
Section \ref{sec:qRac} is devoted to the examples associated with the $q$-Racah polynomials. Two examples corresponding to two different contiguity relations of the $q$-Racah polynomials are studied in detail. Their spectrum is explicitly given. Section \ref{sec:out} discusses different avenues for future research that this work suggests.

\section{XY model and free fermions \label{sec:XY}}

This section aims to fix the notations while recalling the definition of the XY model and the usual methods to find its spectrum.

\paragraph{Inhomogeneous XY model.} The model describes a chain of $N+1$ spins $1/2$ interacting between
nearest neighbors and with an external magnetic field. The strengths of the interaction and of the magnetic field depend on the site.
The quantum Hamiltonian of the inhomogeneous XY model with open boundaries is given by
\begin{align}\label{eq:Ham}
    \cH=\sum_{j=0}^{N-1} \Big(
    (\alpha_j+\gamma_j) \sigma_j^x\sigma_{j+1}^x
    + (\alpha_j-\gamma_j) \sigma_j^y\sigma_{j+1}^y \Big)-\sum_{j=0}^N\beta_j \sigma_j^z\,,
\end{align}
where $\sigma_j^x$, $\sigma_j^y$ and $\sigma_j^z$ are the Pauli matrices:
\begin{align}
    \sigma^x=\begin{pmatrix}
        0&1\\
        1&0
    \end{pmatrix}
    \, ,\quad\sigma^y=\begin{pmatrix}
        0&-i\\
        i&0
    \end{pmatrix}\,,\quad \sigma^z=\begin{pmatrix}
        1&0\\
        0&-1
    \end{pmatrix}\,,
\end{align}
acting on the $j^{th}$ $\CC^2$-space of $(\CC^{2})^{\otimes( N+1)}$ 
and $\alpha_j$, $\beta_j$ and $\gamma_j$ are real parameters.
For $\gamma_j=0$, this model reduces to the XX model. The case when $\alpha_j$, $\beta_j$ and $\gamma_j$ are independent of $j$ is the homogeneous model solved exactly in \cite{lieb1961two} using the Jordan--Wigner transformation \cite{JW}. 
The fact that the model is one-dimensional is important for this transformation to apply.
However, let us remark that there exist some attempts to generalize to
higher dimension \cite{VC,Gal} or to Y-junction \cite{CT}, but the addition of supplementary sites is necessary in these cases.

\paragraph{Jordan--Wigner transformation.} 
 This transformation is based on the introduction of the following operators
\begin{align}
    &c_j=\sigma_0^z\dots \sigma_{j-1}^z \sigma^-_j\,,
    \qquad c_j^\dagger=\sigma_0^z\dots \sigma_{j-1}^z \sigma^+_j\,,
\end{align}
where 
\begin{align}
    \sigma^+=\frac12(\sigma^x+i\sigma^y)=\begin{pmatrix}
        0&1\\
        0&0
    \end{pmatrix}
    \, ,\quad\sigma^-=\frac12(\sigma^x+i\sigma^y)=\begin{pmatrix}
        0&0\\
        1&0
    \end{pmatrix}\,.
\end{align}
These operators are fermionic operators since they satisfy, for $0\leq i,j \leq N$, 
\begin{align}
    \{c_i,c_j\}=0\,,\quad \{c^\dagger_i,c^\dagger_j\}=0\,,\quad \{c^\dagger_i,c_j\}=\delta_{i,j}\,,
\end{align}
where $\{\cdot,\cdot\}$ is the anticommutator and $\delta_{i,j}$ is the Kronecker symbol.
These fermionic operators allow us to rewrite the Hamiltonian \eqref{eq:Ham} as the following free-fermion model:
\begin{align}
    \cH=2\sum_{j=0}^{N-1}\Big( \alpha_j(c_j c_{j+1}^\dagger+  c_{j+1}c_j^\dagger) +\gamma_j(c_j c_{j+1}+ c_{j+1}^\dagger c_j^\dagger)  \Big) +\sum_{j=0}^N \beta_j (2c_j c_{j}^\dagger-1) \,.
\end{align}
It is called a free-fermion model because it is quadratic in the fermionic operators.
This Hamiltonian $\cH$ can be rewritten more compactly as:
\begin{align} \label{eq:Ham2}
    \cH=\bc^\dagger H \bc\,,
\end{align}
where $\bc^\dagger=(c_0,\dots, c_N,c_0^\dagger,\dots c_N^\dagger)$ and $H$ is the $(2N+2)\times (2N+2)$-matrix 
\begin{align}
 H=\left(\begin{matrix}
     A & B \\
     -B & -A
   \end{matrix}\right)\,,
\end{align}
with $A$ a symmetric $(N+1)\times (N+1)$-matrix and $B$ an anti-symmetric $(N+1)\times (N+1)$-matrix taking the following forms
\begin{align}
 A=\left(\begin{matrix}
     \beta_0 & \alpha_0 &  & \\
     \alpha_0 & \beta_1 & \alpha_1 & \\
     & \ddots & \ddots & \ddots\\
     &&\alpha_{N-2} & \beta_{N-1} & \alpha_{N-1}\\
     &&&\alpha_{N-1} & \beta_N
   \end{matrix}\right)\,,\qquad 
   B=\left(\begin{matrix}
     0 & \gamma_0 &  & \\
    - \gamma_0 & 0 & \gamma_1 & \\
     & \ddots & \ddots & \ddots\\
     &&-\gamma_{N-2}& 0 & \gamma_{N-1}\\
     &&&-\gamma_{N-1}& 0
   \end{matrix}\right)\,.
\end{align}
The vector $\bc$ is a column vector with components $c_0^\dagger,\dots, c_N^\dagger,c_0,\dots c_N$.

\paragraph{Properties of $H$.} The matrix $H$ is real symmetric and has real eigenvalues. In addition, using linearity and invariance by permutation of the rows and columns of its characteristic polynomial
\begin{equation}
 \mathcal{P}(x)= \left|\begin{matrix}
     x-A & -B \\
     B & x+A
   \end{matrix}\right|\,,
\end{equation}
one gets $\mathcal{P}(-x)=\mathcal{P}(x)$.
Therefore its diagonalized form can be written as follows 
\begin{align}
 H^{diag}=T^t H T=\left(\begin{matrix}
     \Lambda& 0\\
    0 & -\Lambda
   \end{matrix}\right)\,,
\end{align}
with $T$ an orthogonal transition matrix (\textit{i.e.} $T^tT=1\!\!1_{N+1}$) and $\Lambda$ a $(N+1)\times (N+1)$-diagonal matrix with diagonal entries $\Lambda_0,\Lambda_1,\dots,\Lambda_N$, which are positive real numbers.

The spectral problem associated to the eigenvalue $\Lambda_j$ can be written 
\begin{equation}\label{eq:specj}
 H\left(\begin{array}{c}
         \psi_j\\ \phi_j
        \end{array}
\right)=\Lambda_j \left(\begin{array}{c}
         \psi_j\\ \phi_j
        \end{array}\right)\,,
\end{equation}
where $\psi_j$ and $\phi_j$ are two vectors of $(N+1)$ components.
Using the special form of $H$, one can show that the eigenvector associated to the eigenvalue $(-\Lambda_j)$ is 
\begin{equation}
 H\left(\begin{array}{c}
         \phi_j\\ \psi_j
        \end{array}
\right)=-\Lambda_j \left(\begin{array}{c}
         \phi_j\\ \psi_j
        \end{array}\right)\,.
\end{equation}
The orthogonal transition matrix $T$ takes the following form
\begin{equation}\label{eq:T}
 T=\left(\begin{array}{c c}
         \Psi & \Phi\\ \Phi & \Psi
        \end{array}
\right)\,,
\end{equation}
where the $j^{th}$ column of $\Psi$ (resp. $\Phi$) is $\psi_j$ (resp. $\phi_j$). 

\paragraph{Spectrum of $\cH$.} 
Using the previous transition matrix, one can rewrite the Hamiltonian $\cH$ as follows:
\begin{align}
    \cH= \bc^\dagger\; T T^t\; H\; T T^t\; \bc\,= \,\bd^\dagger\, H^{diag}\, \bd\,,
\end{align}
where $\bd^\dagger=\bc^\dagger T$. With $\bd^\dagger=(d_0,\dots,d_N,d_0^\dagger,\dots,d_N^\dagger)$, one obtains
\begin{align}
    &d_k=\sum_{k=0}^N \psi_{jk} c_j+\sum_{k=0}^N \phi_{jk} c_j^\dagger\,,
    \qquad d_k^\dagger=\sum_{k=0}^N \psi_{jk} c_j^\dagger+\sum_{k=0}^N \phi_{jk} c_j\,.
\end{align}
Using $TT^t=1\!\!1_{N+1}$ and the explicit form \eqref{eq:T} of $T$, one can show  that the previous transformation on fermionic operators induced by $T$ is canonical \textit{i.e.} that $d_j,d_j^\dagger$ are fermionic operators, for $0\leq i,j\leq N$,
\begin{align}
    \{d_i,d_j\}=0\,,\quad \{d^\dagger_i,d^\dagger_j\}=0\,,\quad \{d^\dagger_i,d_j\}=\delta_{i,j}\,.
\end{align}
Finally, the Hamiltonian can be rewritten as
\begin{align}
    \cH= \sum_{j=0}^N \Lambda_j \left( 2d_j d_j^\dagger -1 \right)\,.
\end{align}
Denoting by $|\emptyset \rangle $ the vector such that, for $0\leq j\leq N$,
\begin{align}
    d_j|\emptyset \rangle=0\,,
\end{align}
the eigenvectors of $\cH$ are, for $0\leq \ell\leq N$ and $0\leq \epsilon_1< \dots <\epsilon_\ell \leq N$, 
\begin{align}
    V_{\epsilon_1, \dots ,\epsilon_\ell}=d^\dagger_{\epsilon_1}\dots d^\dagger_{\epsilon_\ell}|\emptyset \rangle\,,
\end{align}
with eigenvalues 
\begin{align}
    E_{\epsilon_1, \dots ,\epsilon_\ell}=2 \sum_{j=1}^\ell \Lambda_{\epsilon_j}
    -\sum_{j=0}^N \Lambda_j \,.
\end{align}
The goal of this paper is to provide parameters $\alpha_j$, $\beta_j$ and $\gamma_j$ such that we can provide an explicit expression for the eigenvalues $\Lambda_j$ and for the transition matrix $T$. 

\section{Exact solution and contiguity relations \label{sec:exact}}

In this section, we show that the entries of the eigenvectors satisfy nice properties recognized as $B_2$-contiguity relations. 

\paragraph{Eigenvectors of $H$.} Building upon the framework established in \cite{lieb1961two}, the spectral problem for $H$ \eqref{eq:specj} can be reformulated:
\begin{subequations}\label{eq:AB1}
\begin{align}
&A \psi_j+B\phi_j=\Lambda_j\psi_j\,,\\
&-B \psi_j-A\phi_j=\Lambda_j\phi_j\,,
\end{align}
\end{subequations}
which are equivalently given by
\begin{subequations}\label{eq:AB+}
    \begin{align}
&(A-B) (\psi_j-\phi_j)=\Lambda_j(\psi_j+\phi_j)\,,\\
&(A+B) (\psi_j+\phi_j)=\Lambda_j(\psi_j-\phi_j)\,.
\end{align}
\end{subequations}
Introducing the following notations for the $k^{th}$ components of the vectors $\psi_j-\phi_j$ and $\psi_j+\phi_j$,
\begin{align}
    (\psi_j-\phi_j)_k=P_k(j)\,,\qquad  (\psi_j+\phi_j)_k=Q_k(j)\,,
\end{align}
equations \eqref{eq:AB+} become, for $0\leq k\leq N$,
\begin{subequations}\label{eq:recuPQ}
    \begin{align}
  &  \beta_k P_k(j) +(\alpha_k-\gamma_k) P_{k+1}(j)+(\alpha_{k-1}+\gamma_{k-1}) P_{k-1}(j)=\Lambda_j Q_k(j)\,,\\
   &  \beta_k Q_k(j) +(\alpha_k+\gamma_k) Q_{k+1}(j)+(\alpha_{k-1}-\gamma_{k-1}) Q_{k-1}(j)=\Lambda_j P_k(j)\,,
\end{align}
\end{subequations}
with the conventions $\alpha_{-1}=\gamma_{-1}=\alpha_{N}=\gamma_{N}=0$.
An important observation at this point is that these last relations bear a similarity to the $B_2$-contiguity relations studied in the context of orthogonal polynomials.

\paragraph{$B_2$-contiguity relations.} 
We denote by $\{ R_i(x; \rr)\}_{0\leq i\leq N}$ a discrete family of polynomials of the ($q$-)Askey scheme where $\rr$ is the list containing all the parameters of these polynomials (in particular, it includes the integer $N$).

The $B_2$-contiguity relations relations take the following form
\begin{subequations}\label{eq:lpm}
\begin{align}\label{eq:lp}
&\lambda^{+}_{x; \rr}\; R_i(x; \rr)= \Phi^{+1,+}_{i}\ R_{i+1}(\ox; \orr)+\Phi^{0,+}_{i}\ R_{i}(\ox; \orr)+\Phi^{-1,+}_{i}\ R_{i-1}(\ox; \orr)\,,\\
\label{eq:lm}
&\lambda^{-}_{x; \rr}\; R_i(\ox; \orr)= \Phi^{+1,-}_{i}\ R_{i+1}(x; \rr)+\Phi^{0,-}_{i}\ R_{i}(x; \rr)+\Phi^{-1,-}_{i}\ R_{i-1}(x; \rr)\,,
\end{align} 
\end{subequations}
where $\ox$ is a linear transformation of $x$ and $\orr$ is the list of modified parameters in $\rr$. The functions $\lambda^{\pm}_{x; \rr},\;\Phi^{+1,\pm}_{i},\;\Phi^{0,\pm}_{i}$ and $\Phi^{-1,\pm}_{i}$ that appear in the previous relations will be given explicitly in the following for the $q$-Racah polynomials.

Let normalize these polynomials as follows, for $0\leq i\leq N$,
\begin{align}
  P_i(x)=\sqrt{\lambda^{+}_{x; \rr}
  \Phi_0^{0,-} \prod_{k=0}^{i-1}\frac{\Phi_k^{0,+}\Phi_k^{+1,-}}{\Phi_k^{0,-}\Phi_{k+1}^{-1,+}}} \;R_i(x; \rr)\,,\qquad
  Q_i(x)=\sqrt{\lambda^{-}_{x; \rr}
    \Phi_0^{0,+}\prod_{k=0}^{i-1}\frac{\Phi_k^{0,-}\Phi_k^{+1,+}}{\Phi_k^{0,+}\Phi_{k+1}^{-1,-}}}\; R_i(\ox; \orr)\,.
\end{align}
The previous functions $P_i(x)$ and $Q_i(x)$ satisfy relations \eqref{eq:recuPQ} if $\oN=N$ and if the following constraints are satisfied, for $0\leq i \leq N$,
\begin{align}\label{eq:comsPhi}
    \frac{\Phi_i^{0,-}\Phi_{i}^{+1,+}\Phi_{i+1}^{-1,+}\Phi_{i+1}^{0,-}}{\Phi_{i}^{0,+}\Phi_{i}^{+1,-}\Phi_{i+1}^{-1,-}\Phi_{i+1}^{0,+}}=1\;.
\end{align}
The normalizations for $R_i(x;\rr)$ and $R_i(x;\orr)$ have been chosen such that the two different eigenvalues in \eqref{eq:lpm} become equal. 
The coefficients in \eqref{eq:recuPQ} read
\begin{equation}\label{eq:coeffsex}
    \Lambda_j=\sqrt{\lambda^{+}_{j; \rr}\lambda^{-}_{j; \rr}}\,,\quad \beta_j=\sqrt{\Phi_{j}^{0,+}\Phi_j^{0,-}},\quad
2\alpha_j=\sqrt{\Phi_{j+1}^{-1,+}\Phi_{j}^{+1,-}}+\sqrt{\Phi_{j+1}^{-1,-}\Phi_i^{+1,+}},
\end{equation}
\[ 2\gamma_j=\sqrt{\Phi_{j+1}^{-1,-}\Phi_{j}^{+1,+}}-\sqrt{\Phi_{j+1}^{-1,+}\Phi_i^{+1,-}}.\]

In \cite{Contiguity25}, the $B_2$-contiguity relations
for the finite families of orthogonal polynomials in the Askey scheme have been classified. We recall and use these results in Section \ref{sec:qRac} to obtain exactly solvable XY model.

\section{$q$-Racah polynomials and XY model \label{sec:qRac}}

The $q$-Racah polynomials are the most general orthogonal polynomials of the finite families of the $q$-Askey scheme. They are defined by, for $i=0,1,\dots,N$,
\begin{align}\label{eq:qRacah}
R_i(x;\rr)={}_4\phi_3 \left({{q^{-i},\; a b q^{i+1}, \;q^{-x},\;c  q^{x-N}}\atop
{a q,\; b c q,\;q^{-N} }}\;\Bigg\vert \; q;q\right)\,,
\end{align}
where $\rr=a,b,c,N,q$ and $N$ is a non-negative integer. The standard definition of the generalized $q$-hypergeometric function \cite{GasperRahman2004,Koekoek2010} is used:
\begin{align}
{}_{4}\phi_3 \left({{q^{-i},\; a_1,\; a_2,\; a_3 }\atop
{ b_1,\; b_2,\; b_3}}\;\Bigg\vert \; q;z\right)=
\sum_{k=0}^i \frac{(q^{-i},a_1,a_2,a_3;q)_k}{(q,b_1,b_2,b_3;q)_k}  z^k\,,
\end{align}
where
\begin{align}
(b_1,\dots,b_p;q)_k=(b_1;q)_k\dots (b_p;q)_k\,,\qquad (b_i;q)_k=\prod_{\ell=0}^{k-1}(1-b_i q^\ell)\;.
\end{align}
The $q$-Racah polynomials $R_i(x;\rr)$ are polynomials of degree $i$ with respect to the variable $\lambda_{x,\rr}=-(1-q^{-x})(1-c q^{x-N})$.

The $B_2$-contiguity relations for the $q$-Racah polynomials has been classified in \cite{Contiguity25} (a complete list can be found in appendix A of this paper). Focusing on those relations such that $\oN=N$, we see that they all satisfy the constraints \eqref{eq:comsPhi}. In the following, we provide the various exactly solvable inhomogeneous XY models obtained by this procedure.

\subsection{Model based on (qRI/III) and (qRIII/I) contiguity relations}

Let relation \eqref{eq:lp} be the contiguity relation (qRI/III) given in Appendix A of \cite{Contiguity25}. The transformations of the parameters are given by
\begin{align}
    \ox=x +1, \quad \oa=a/q,\quad \ob=b q,\quad \oc=c/q^2\,, \quad\oN=N\,,
\end{align}
and the coefficients are: 
\begin{align*}
  &  \Phi_i^{+1,+}=-\frac{q^{i}(1-q^{i-N})(1-ab q^{i+1})}
  {(1-ab q^{2i+1})(1-a b q^{2i+2})}\,,
 \qquad  \Phi_i^{-1,+}=-\frac{q^{i-N-1}(1-q^i)(1-ab q^{N+i+1})}{(1-ab q^{2i})(1-a b q^{2i+1})}\,,
    \\
    & \Phi_i^{0+}=-\Phi_i^{+1,+}-\Phi_i^{-1,+}\,,\qquad \lambda_{x,\rr}^+=\frac{(1- q^{-x-1})(1-c q^{x-N-1})}{(1-bc )(1-a)}
 \,.
\end{align*}
The associated relation \eqref{eq:lm} is the contiguity relation (qRIII/I) given in Appendix A of \cite{Contiguity25} for which the parameters are shifted.
The coefficients read
\begin{align*}
 &  \lambda_{x,\rr}^-=(1-c q^{x})(1-q^{N-x})\,,
 \\
  &  \Phi_i^{+1,-}=\frac{(1-q^{N-i})(1-a q^{i})(1-a q^{i+1})(1-ab q^{i+1})(1-bc q^{i})(1-bc q^{i+1})}{(1-a)(1-bc )(1-ab q^{2i+1})(1-ab q^{2i+2})}\,,
  \\
  &  \Phi_i^{0,-}=\frac{(1-a q^{i})(1-b q^{i+1})(1-bc q^{i})(a q^{i+1}-c )}{q(1-a)(1-bc )(1-ab q^{2i+1})}\left(
  \frac{q(1-q^{N-i})(1-a b q^{i+1})}{1-a b q^{2i+2}}
  +\frac{(1-q^{-i})(1-ab q^{N+i+1})}{1-ab q^{2i}}
  \right)\,,
  \\
    &  \Phi_i^{-1,-}=\frac{(1-q^{-i})(a q^{i}-c )(a q^{i+1}-c)(1-b q^{i})(1-b q^{i+1})(1-ab q^{N+i+1})}{q(1-a )(1-bc )(1-ab q^{2i})(1-ab q^{2i+1})}\,.
\end{align*}
By direct computations, we show that the constraints \eqref{eq:comsPhi} are satisfied with these previous coefficients. Therefore, the XY model with the coefficients $\alpha_j$, $\beta_j$ and $\gamma_j$ given by \eqref{eq:coeffsex}
is exactly solvable with the eigenvalues 
\begin{align}
    \Lambda_j&=\sqrt{\frac{(1-c q^{j})(1-q^{N-j})(1- q^{-j-1})(1-c q^{j-N-1})}{(1-bc )(1-a)}}\,.
\end{align}

\subsection{Model based on (qRII/IV) and (qRIV/II) contiguity relations}

Let relation \eqref{eq:lp} be the contiguity relation (qRII/IV) given in Appendix A of \cite{Contiguity25}. The transformations of the parameters are given by
\begin{align}
   \ox=x, \quad \oa=a/q,\quad \ob=b q,\quad \oc=c\,,\quad \oN=N\,,
\end{align}
and the coefficients are: 
\begin{align*}
  &  \Phi_i^{+1,+}=-\frac{(q^N-q^i)(1-ab q^{i+1})(1-bc q^{i+1})(1-bc q^{i+2})}{(1-bc q)(1-ab  q^{2i+1})(1-ab  q^{2i+2})}\,,
  \\
    &  \Phi_i^{-1,+}=-\frac{b q(1-q^i)(c-a q^{i-1})(c-a q^i)(1-ab q^{N+i+1})}{a(1-bc q)(1-ab q^{2i})(1-ab q^{2i+1})}\,,
    \\
      &  \Phi_i^{0+}=\lambda_{0,\rr}^+-\Phi_i^{+1,+}-\Phi_i^{-1,+}\,,\qquad  \lambda_{x,\rr}^+=\frac{(1-a q^{x})(c-a q^{N-x})}{a(1-a)}\,.
\end{align*}
The associated relation \eqref{eq:lm} is the contiguity relation (qRIV/II) given in Appendix A of \cite{Contiguity25} for which the parameters are shifted.
The coefficients read
\begin{align*}
  &  \Phi_i^{+1,-}=\frac{(1-a q^{i})(1-a q^{i+1})(1-ab q^{i+1})(q^i-q^N)}{(1-a )(1-ab q^{2i+1})(1-ab q^{2i+2})}\,,
   \Phi_i^{-1,-}=-\frac{a (1-q^i)(1-b q^{i})(1-b q^{i+1})(1-ab q^{N+i+1})}{bq(1-a )(1-a b q^{2i})(1-ab q^{2i+1})}
    \\
      &\lambda_{x,\rr}^-=\frac{(1-bc q^{x+1})(1-b q^{N-x+1})}{b q(1-b cq)}\,,\qquad  \Phi_i^{0-}=\lambda_{0,\rr}^+-\Phi_i^{+1,+}-\Phi_i^{-1,+}\,.
\end{align*}
By direct computations, we show that the constraints \eqref{eq:comsPhi} are satisfied with these previous coefficients. Therefore, the XY model with the coefficients $\alpha_j$, $\beta_j$ and $\gamma_j$ given by \eqref{eq:coeffsex}
is exactly solvable with the eigenvalues 
\begin{align}
    \Lambda_j&=\sqrt{\frac{(1-a q^{j})(c-a q^{N-j})(1-bc q^{j+1})(1-b q^{N-j+1})}{ab q(1-a)(1-b cq)}}\,.
\end{align}

\section{Outlooks \label{sec:out}}

This paper establishes a novel connection between exactly solvable quantum models and special functions, paving the way for several interesting research avenues.

The models introduced herein depend on the parameters $a,b,c$ and $q$. It would be highly valuable to explore different limits of these parameters to connect these new models with other orthogonal polynomials within the ($q$-)Askey scheme. Such explorations are expected to reveal diverse behaviors of the associated eigenvalues. 

We focused solely on the eigenvalues to demonstrate the feasibility of obtaining exact results for inhomogeneous XY spin chains. However, since the eigenvectors are associated with orthogonal polynomials, it should be possible to compute other physical quantities, such as density, correlation functions, phase diagram and entanglement entropy. For the last quantity, we wonder if the computation involving Heun operators is also possible as for the XX model \cite{crampe2019free}.

Another natural extension of this work involves generalizing the proposed construction. In reference \cite{Contiguity25}, it was demonstrated that the $B_2$-contiguity relations correspond to spectral transforms of polynomials, specifically Christoffel and Geronimus transforms. Exploring more general Christoffel and Geronimus transforms could potentially yield new exactly solvable XY models.

Finally, the potential role of matrix valued orthogonal polynomials warrants depeer investigation. Notably, the system described in equation \eqref{eq:recuPQ} can be reformulated as a three term recurrence relation with matrix coefficients, a structure
closely tied to the theory of matrix or vector valued orthogonal polynomials. The connection between integrable systems and such polynomials has been investigated, primarily in the development of non-Abelian generalizations of classical integrable structures, such as the Painlevé equations and the Toda lattice, see for instance \cite{DMR,Cafasso,Ariznabarreta}. Consequently, a systematic exploration of this matrix polynomial structure
may yield new insights for the studies of XY models.

\paragraph{Acknowledgements:} The authors thank warmly G.~Parez for primary discussions about this work.
P.-A.~Bernard holds an Alexander-Graham-Bell
scholarship from the Natural Sciences and Engineering Research Council (NSERC) of Canada. N.~Cramp\'e is partially supported by the international research project AAPT of the CNRS. L.~Vinet is funded in part by a Discovery Grant from the Natural Sciences and Engineering Research Council (NSERC) of Canada. Q. Labriet and L. Morey enjoy postdoctoral fellowships provided by this grant.

\bibliographystyle{utphys}
\bibliography{XYinh}

\end{document}